\documentclass[a4paper,11pt]{article}
\pdfoutput=1
\usepackage{jheppub}
\usepackage{graphicx}
\usepackage{amsmath}
\usepackage{amsfonts}
\usepackage{amssymb}
\usepackage{slashed}
\usepackage{diagbox}
\usepackage[colorlinks=true,linkcolor=blue]{hyperref}
\usepackage{bm}
\usepackage{empheq}
\usepackage[super]{nth}
\usepackage{physics}

\def\tA{\theta_{\mathrm{a}}}

\def\half{\frac{1}{2}}

\def\LL{\mathcal{L}}
\def\be{\begin{equation}}
\def\ee{\end{equation}}
\def\bea{\begin{eqnarray}}
\def\eea{\end{eqnarray}}

\def\Eq#1{Eq.~\eqref{#1}}

\def\nax{n_{\mathrm{ax}}}
\def\naxbar{\bar{n}_{\mathrm{ax}}}

\def\nmissbar{\bar{n}_{\mathrm{misalign}}}
\def\atan{\mathrm{atan}}

\def\ma{m_a}
\def\mac{m_{a,\mathrm{c}}}
\def\fa{f_a}

\def\tauo{\tau_*}
\DeclareMathOperator{\tStep}{a_\tau}
\DeclareMathOperator{\xStep}{a_x}
\DeclareMathOperator{\tStepsq}{a_\tau^2}
\DeclareMathOperator{\xStepsq}{a_x^2}

\title{Misalignment vs Topology in Axion-Like Models}
\author{Aidan Chaumet,}
\author{Guy D.\ Moore}

\affiliation{Institut f\"ur Kernphysik, Technische Universit\"at Darmstadt\\
Schlossgartenstra{\ss}e 2, D-64289 Darmstadt, Germany}
\emailAdd{achaumet@theorie.ikp.physik.tu-darmstadt.de}
\emailAdd{guy.moore@physik.tu-darmstadt.de}

\abstract{
  Cosmological axions, an important dark matter candidate, are
  generated when a scalar field with a spontaneously broken $O(2)$
  invariance develops an explicitly
  $O(2)$-breaking tilt cosmologically, leading to coherent production of
  long-wavelength modes which could be dark matter.  It appears that
  the production efficiency of this mechanism is low, despite possible
  production from topological structures.  To understand this better,
  we examine production in $O(N)$ generalizations.  In particular we
  highlight the importance of how fast the explicit $O(N)$-breaking potential
  turns on.
}

\begin{document}
\maketitle
\section{Introduction}
\label{sec:intro}

The QCD axion \cite{Weinberg:1977ma,Wilczek:1977pj}
is a hypothetical particle which is effectively the angular degree of
freedom of a scalar field with a spontaneously broken $O(2)$ symmetry,
when there is also an extremely small explicit breaking of the
symmetry (making the axion a pseudo-Goldstone boson).
It represents a nearly ideal dark matter candidate
\cite{Preskill:1982cy,Abbott:1982af,Dine:1982ah}.
One scenario for a cosmological abundance of axions to develop
\cite{Davis:1986xc}, recently argued to be the most likely to occur
\cite{Visinelli:2009zm,Visinelli:2014twa}, is that
the $O(2)$ symmetry is restored  during or after inflation, and then
spontaneously breaks early in the Universe's history, leading to a
network of axionic strings which disappear when the explicit breaking
becomes important.  A key feature is that the explicit breaking, due to
anomalous couplings to QCD, ``turns on'' rather abruptly as the
Universe cools; in the relevant temperature range around 1 GeV,
we expect the axion mass to scale as
$m_a^2 \propto T^{-7.6}$
\cite{Wantz:2009mi,Borsanyi:2016ksw}.

Many authors have investigated the production of axions under
these circumstances, both analytically
\cite{Harari:1987ht,Hagmann:1998me,Battye:1993jv,Battye:1994au},
and by using classical field theory simulations on a spacetime lattice
\cite{Yamaguchi:1998gx,Yamaguchi:1999yp,Hiramatsu:2010yu,%
  Hiramatsu:2012gg,axion1,axion2}.
The most reliable approach is to follow the full dynamics of the Axion
field and string network as it breaks up and to count the total axions
at the end, further improving the reliability of the simulation by
incorporating additional physics which accounts for the large string
tension associated with short-distance effects along string cores
\cite{axion4}.  These simulations, which are supported by other
recent large-scale simulations
\cite{Vaquero:2018tib,Buschmann:2019icd},
indicate surprisingly that the total axion production is actually less
than the expected \textsl{baseline} production efficiency, the
so-called misalignment value.%
\footnote{The issue of the correct handling of string cores is
  delicate, and not all recent literature agrees with this result, see
  in particular \cite{Gorghetto:2018myk,Gorghetto:2020qws}.
  Note however that these papers do not follow the axion dynamics
  through the explicit-breaking epoch and count final axions.}
Here the misalignment value is the
number of axions which would be produced if the Universe consisted of
many independent regions, each with an independent, random, and
homogeneous starting value on the vacuum manifold of the $O(2)$ model
before the explicit breaking became important.

Many people, including these authors, find this result bizarre and
counterintuitive.  A field with large inhomogeneities, including
highly energetic topological defects, somehow generates \textsl{fewer}
excitations than it would if it were locally uniform.  We want to
understand this result better by seeing whether something similar
happens in closely related models.  In this paper we will investigate
what happens if, rather than a two-component scalar field with
spontaneously and also very weakly explicitly broken $O(2)$
invariance, we have an $O(N)$
field with $N=3,4,5$.  The case $N=3$ also contains topological
defects -- global monopoles \cite{Kibble:1976sj} -- and $N=4$ has
textures \cite{Turok:1989ai}.  But field-theoretical simulations of
these models are more reliable than for the case of $O(2)$, because the
energies associated with these topological objects are not dominated
by their cores but are instead dominated by longer-range field
gradients which should be correctly described by a lattice
implementation.  Therefore the small lattice-spacing limit can be
addressed robustly, and our results should be secure.

In Section \ref{sec:model} we will set up the scenario we consider.
Then Section \ref{sec:numerics} describes our numerical implementation
of $O(N)$ theory with spontaneous and weak explicit breaking on the
lattice, from the initial conditions through to counting axions at the
end.  Section \ref{sec:results} presents our results, and we end with
a discussion.  But here we will give a very brief precis of our
results.  The production of angular (axion-like) excitations is highly
sensitive to how suddenly the explicit symmetry breaking turns on with
time.  If the explicit breaking turns on very abruptly, the field
gradients are not very important and the misalignment estimate is
fairly accurate.  If the breaking turns on slowly, for instance, if
the physical (as opposed to conformal) mass is constant, then theories
with many field components produce substantially more angular
excitations than in the misalignment mechanism.  Therefore the very
abrupt turn-on of the axion mass in cosmology may be key to
understanding why the resulting axion abundance is so modest.

\section{Axion-like models}

\label{sec:model}
We will describe the axion model and its generalization to
$N$ components first by expressing the model, then by outlining its
cosmological evolution, and then by showing how nontrivial
topological structures play a role in the evolution, and why they are
under better control for the case $N \geq 3$ than for $N=2$.

\subsection{Field content}

For the purposes of cosmology, an axion field can be summarized as a
two-component scalar field model $\Phi = (\Phi_1,\Phi_2)$ with an
$O(2)$ symmetry which is spontaneously broken and also very weakly
explicitly broken.  The Lagrangian density%
\footnote{We use a metric with $[{-}{+}{+}{+}]$ signature.}
is
\be
\label{eqn:standardAxionLagr}
- \LL_{\mathrm{axion}} =\half \sum_{i=1}^2 \partial^\mu\Phi_i\partial_\mu\Phi_i
 + \frac{\lambda}{8}\left(\sum_{i=1}^2 \Phi_i^2 - \fa^2 \right)^2
 - \ma^2\fa\Phi_1,
\ee
where the $\Phi_i$ are two real fields and the choice of explicit
symmetry breaking in the $\Phi_1$ direction is an arbitrary choice.
Temporarily considering the case without explicit symmetry breaking,
$\ma^2 = 0$, the spontaneous-symmetry-breaking potential selects a
vacuum manifold with $\sqrt{\Phi_1^2+\Phi_2^2}=f_a$,
and we can bring out the physics by writing in terms of radial
and angular variables:
$\Phi_i = (|\Phi| \cos(\tA),|\Phi| \sin(\tA))$,
with $|\Phi| = \sqrt{\Phi_1^2+\Phi_2^2}$ and $\tA = \atan(\Phi_2/\Phi_1)$.
The minimum of the axion potential is then achieved whenever $|\Phi| = \fa$,
and the free choice of $\tA$ parametrizes the vacuum manifold.
We then distinguish radial excitations
$|\Phi| = \fa + h$, with $h$ the radial (Higgs or saxion) fluctuation,
and angular or axion excitations $a=\fa\, \tA$.
The radial mass is given by $m_r^2 = \lambda\fa^2$
while the axions are massless before introducing explicit symmetry
breaking.  When explicit symmetry breaking is present, the global
potential minimum lies at approximately $(h,a)=(0,0)$ and the axion
mass is given by $m_a^2$.
The value of the symmetry breaking scale can be constrained from
cosmological observations, and in the physically interesting regime
there is a clear separation of scales given by radial and axion mass,
$m_r \sim 10^{30} m_a$ such that we do not need to consider radial
dynamics \cite{Turner:1989vc,Raffelt:1990yz,Raffelt:1999tx} except in
the cores of topological defects (see below).
In a cosmological setting the angular excitations will have
wavelengths of order of the Hubble scale and amplitudes of order 1
radian, representing quantum occupancies of order
$\fa^2/H^2$.  For typical applications this is $\mathcal{O}(10^{60})$,
making a classical field approximation very well justified.

This model can immediately be generalized to $N$-component scalar
fields with spontaneously broken $O(N)$ invariance plus a very weak
explicit breaking.  The Lagrangian density is
\be
\label{eqn:generalAxionLagr}
- \LL = \half \sum_{i=1}^N
 \partial^\mu\Phi_i\partial_\mu\Phi_i
 + \frac{\lambda}{8}\left(\sum_{i=1}^N \Phi_i^2 - \fa^2 \right)^2
 - \ma^2\fa\Phi_1.
\ee
Instead of a single angular variable $\tA$ we now have $N-1$ angular
variables corresponding to the $N-1$ Goldstone modes of spontaneous
$O(N) \to O(N-1)$ breaking.  When $\ma^2 \neq 0$ the true minimum
resides at $\Phi_1=\fa$ and $\Phi_{2,\ldots,N} = 0$, the angular 
excitations correspond at linearized order to the $\Phi_{2,\ldots N}$
excitations, and they all have mass-squared values equal to $\ma^2$.

An important feature of the axion model in cosmology is that the axion
mass squared $\ma^2$ is \textsl{not} directly a Lagrangian mass, but
is rather an effect due to nontrivial interactions with the QCD
sector.  As such it is strongly temperature dependent above
$T = 200\:\mathrm{MeV}$, which is the regime where the most
interesting dynamics will occur.  Therefore we are interested in cases
where $\ma^2$ shows a strong temperature dependence.  Even though we
are not aware of cases where this is relevant for an $O(N)$ model, we
will consider similar temperature dependence when we generalize to
these models, because our goal is to understand how this among other
things affects the final particle number generation.

\subsection{Cosmological evolution}

Next we consider how the $O(2)$ model and its $O(N)$ generalization
would behave cosmologically, with a focus on the final excitation
density.
If symmetry is restored at some point in the early Universe and then
breaks spontaneously, different space regions will rapidly reach the
vacuum manifold $|\Phi| = \fa$ but will randomly select
different angles $\tA$.  Subsequent (dissipative) dynamics tend to
align the value of $\tA$ over causally connected regions of space, in
order to minimize the gradient energy;
the field undergoes \textsl{local ordering dynamics}.  However,
causality prevents this ordering from occurring globally; two points
with non-overlapping past light cones cannot have correlated field
values.

We are concerned with evolution in the radiation-dominated early
universe, which features a time-varying scale factor $a$
with a Hubble scale $H=\dot{a}/a = 1/2t$.
We will work in comoving coordinates $x_i$ and conformal time $\tau$,
so the spacetime metric is
\begin{equation}
  \label{gmunu}
  g_{\mu\nu} = \left( \frac{\tau}{\tau_0} \right)^2 \eta_{\mu\nu} \,,
  \quad
  \eta_{\mu\nu} = \mathrm{Diag}[-1,1,1,1] \,.
\end{equation}
The Lagrangian density is then
\begin{equation}
  \label{Lconformal}
  -\sqrt{-g} \: \LL = \tau^2 \left( \half \sum_i \eta^{\mu\nu}
  \partial_\mu \Phi_i \partial_\nu \Phi_i
  + \frac{\lambda \tau^2}{8\tau_0^2}
  \left( \sum_i \Phi_i^2 - \fa^2 \right)^2
  - \frac{\tau^2 \ma^2 \fa}{\tau_0^2} \Phi_i \right).
\end{equation}
The explicit factor of $\tau^2$ in front will give rise to Hubble
damping, while the extra $\tau^2$ factor attached to $\ma^2$ makes the
mass more time dependent in these conformal-time coordinates.
We will call the combination $\tau^2 \ma^2/\tau_0^2 \equiv \mac^2$ in
what follows.  In addition, $\ma^2$ will have explicit temperature,
and therefore time, dependence.
Within the axion model this arises because 
$\ma^2$ arises from a coupling to QCD dynamics which is temperature
dependent; we have (see for instance \cite{diCortona:2015ldu})
$\ma^2 = \chi(T) / \fa^2$ with $\chi(T)$ the QCD
topological susceptibility.  Instanton gas estimates suggest
$\ma^2 \propto T^{-6.7} \propto \tau^{6.7}$
\cite{Wantz:2009it}
but more recent lattice studies find a still stronger dependence,
$\ma^2 \propto T^{-7.6} \propto \tau^{7.6}$
\cite{Borsanyi:2016ksw}.  We will parametrize this with a coefficient
$n$: $\ma^2 \propto T^{-n} \propto \tau^n$, with fixed physical mass
being $n=0$ and the axion case being close to $n=7.6$; we will use
$n=7$ to represent this case in what follows, and we will contrast the
field behavior in these two cases, $n=0$ ($\mac^2 \propto \tau^2$)
and $n=7$ ($\mac^2 \propto \tau^9$).

So long as $\mac^2 \tau^2 < 1$, there is no time for the explicit symmetry
breaking to play a role in the dynamics.  The field takes values all
over its vacuum manifold, the value varies with a spatial coherence
length comparable to the system's age, and there is no useful
definition of angular particle number.
However, $\mac^2 \tau^2 \propto \tau^{n+4}$ and one
rather abruptly enters the opposite regime where the explicit symmetry
breaking is important.  Once this happens, the fields are drawn to the
unique minimum of the potential.  After some complicated dynamics, at
late times the field will settle near the global minimum, but with
residual spatially inhomogeneous small-angle oscillations.

Once these oscillations truly become small,
$\Delta \Phi_i^2 \ll \fa^2$, one can define a particle number which is
an adiabatic invariant in the dual small-angle and
$\mac^2 \tau^2 \gg 1$ approximations.  In terms of the Fourier
spectrum of the fields $\Phi_{i}(k)$, this invariant is
\be
\label{naxdef}
\nax = \frac{\tau^2}{2} \int \frac{\dd[3]{k}}{(2\pi)^3}
\left( \sqrt{\mac^2+k^2} \sum_{i=2}^N \Phi^2_i(k)
+ \frac{\sum_{i=2}^N \dot\Phi^2_i(k)}{\sqrt{\mac^2+k^2}} \right)
\ee
where the factor $\tau^2$ in front counteracts the loss of particle
number due to Hubble damping.  In standard space and time coordinates
the corresponding particle number (without the $\tau^2$ factor)
diminishes as $a^{-3}$, staying fixed per comoving volume, and
corresponds to the axion number density in the universe.
Our goal will be to determine $\nax$ and to understand how it depends
on the number of field components.

\subsection{The misalignment estimate}

\begin{figure}
\centering
  \includegraphics[width = 0.8\textwidth]{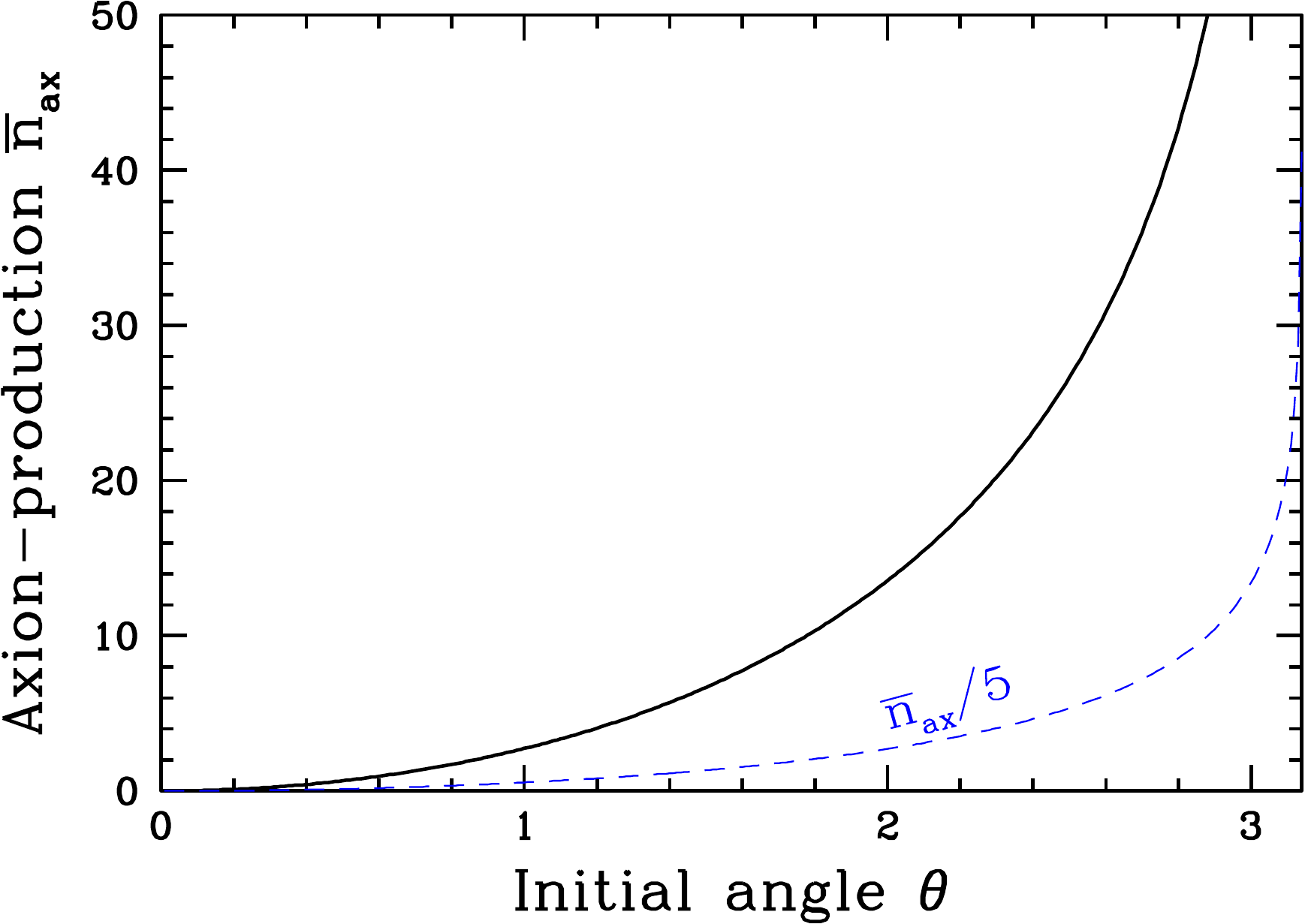}
  \caption{How the axion production efficiency depends on the
    starting misalignment angle.  The curve is quadratic for small
    angles, but rises towards a logarithmic divergence
    at $\tA=\pi$.}
  \label{fig:misalign1}
\end{figure}
It is useful to begin with a very crude estimate for the final axion
density.  Consider what dynamics would occur if the field
were spatially homogeneous.  Consider first the 2-component field.
In this case $\tA$ starts with some initial value $\tA \in [-\pi,\pi]$
and vanishing time derivative (due to Hubble damping).
Neglecting gradient terms in the equations of motion,
the field evolves according to:
\be
\label{eqn:mislagnOde}
\partial_\tau^2 \tA + \frac{2}{\tau} \partial_\tau \tA
+ \mac^2\sin(\tA) = 0.
\ee
Numerically integrating this equation, one rather easily finds the
final (comoving) axion density for each initial $\tA$ value:
$\nax(\tA)$.  We plot the $\tA$ dependence, for $n=7.6$,
in Figure \ref{fig:misalign1}.
The misalignment estimate is to replace the true field dynamics with
random initial conditions with the above dynamics, but
average the final axion production uniformly over the range of
possible initial $\tA$ values:
$\nax = \frac{1}{\pi} \int_0^\pi \nax(\tA) \dd{\tA}$.
This does not correspond to the axion
production in any physical scenario, but it sets a reasonable baseline
expectation for axion production for the random phase case.

The generalization to $N$ field components is straightforward.
The field begins somewhere on $S^{N-1}$ the $N-1$-sphere.
But we can perform an $SO(N-1)$ rotation about the $\Phi_1$ axis
to place the field excitation entirely in the
$(\Phi_1,\Phi_2)$ plane.  Since the initial time derivatives are
zero and $SO(N-1)$ invariance remains unbroken, the field values will
stay in this plane and the dynamics are identical to the two-component
case.  All that changes is the weighting of the initial angle:
we now have
\be
\label{naxmisalign}
\nax[SO(N)] = \frac{\int_0^\pi \nax(\tA) \sin^{N-2}(\tA) \dd{\tA}}
    {\int_0^\pi \sin^{N-2}(\tA) \dd{\tA}} \,.
\ee
This weights the integral more heavily near $\tA=\pi/2$.  Since the
function is concave, the larger $N$ is, the fewer axions are
predicted.  More field components lead to fewer excitations!

\subsection{Topological objects and their role}

We remarked already that the detailed dynamics of field ordering can
be complex.  One thing which makes them particularly complex is the
presence of topological objects, at least for $N=2,3,4$.
Here we will give a very quick review; the topic is well addressed in
the existing literature
\cite{Kibble:1976sj,Vilenkin:1981sd,Vilenkin:1982ks,Turok:1989ai,
Bennett:1990xy,Preskill:1992ck}.

We already remarked that the field evolution and the presence of
gradient energies leads to locally smooth but globally random fields
during the early dynamics, $\mac \tau \leq 1$.
Such a field configuration
generically features topological defects \cite{Kibble:1976sj}.
For an $N$-component field, the vacuum manifold is $S^{N-1}$. Removing
a $3-N$ dimensional curve from space -- or a $4-N$ dimensional curve
from spacetime -- leaves a region where the field can vary
nontrivially around the vacuum manifold.  For $N=2$ one removes a line
(or curve) and the field can wind around the circle as one goes around
the line.  For $N=3$ one removes a point, and the spherical surface
enclosing the point can have the field direction on $S^2$ wrap
nontrivially.  In each case, the field must leave the vacuum manifold
very close to the removed line/point; the size of this ``core'' is of
order the inverse radial mass $m_r^{-1}$.  Since the physical case
corresponds to $H/m_r \sim 10^{-30}$ but on the lattice we can only
obtain $H/m_r \sim 10^3$, the size of this core region is exaggerated
on the lattice.

To assess the importance of the mistreatment of the defect core size,
we need to estimate how large a role the core plays in the defect's
energetics. The field gradient at a distance $r$ from the defect core
is of order $|\nabla \Phi| \sim \fa/r$, leading to an energy which
scales as
\begin{align}
  \label{core_energy}
  E & \sim \int_{m_r^{-1}}^{H^{-1}} r dr \frac{\fa^2}{r^2} \propto
  \ln(m_r/H) \,, & N=2 &\; \mbox{(string)}
  \\ \nonumber
  E & \sim \int_{m_r^{-1}}^{H^{-1}} r^2 dr \frac{\fa^2}{r^2} \propto
  H^{-1} -  m_r^{-1} \,, & N=3 &\; \mbox{(monopole)}.
\end{align}
For $N=2$, the short-distance behavior near the string plays a major
role in the energetics and the defect energy scales logarithmically
with $m_r$; but for $N=3$, the short-distance behavior plays a minor
role, with the core energy scaling with $m_r^{-1}$.  Therefore, using
an unphysically small $m_r$ value could lead to logarithmically large
errors for $N=2$, but the errors will be suppressed by $m_r^{-1}$ for
$N=3$ and higher.  This is why we expect the lattice treatment of
$N \geq 3$ to be reliable up to power-suppressed effects, unlike in
the case $N=2$ where it is necessary to turn to effective descriptions
to capture the relevant string-core physics
\cite{axion3,axion4}.

After explicit symmetry breaking becomes relevant, the defects are no
longer strictly topological, since the field now has a unique global
minimum.  For $N=2$ each string is attached to a domain wall with
tension $\sigma = 8 \fa^2 m_a$.  The domain wall is locally stable
because the gradient energy scales as the inverse wall thickness while
the potential energy is linear in the wall thickness.  The wall exerts
a force-per-length on the attached string, which pulls the string
network shut.  The associated dynamics play a large role in the energy
budget.  In contrast, for $N=3$ each monopole is attached to a
``string'' in whose core $\Phi_1 = -\fa$.  However such a string is
\textsl{not locally stable}; the gradient energy is independent of the
string's thickness while the potential energy grows with the string
thickness, so the string tends to collapse to zero thickness and
fragment.  Therefore the monopole network can annihilate without the
monopoles needing to physically travel to reach each other, and again
the finite $m_r$ value does not play much of a role in the system's
energetics.  The conclusion is that, for $N \geq 3$, topological
structures can play a role but they are well described provided that
the ratios $m_r/m_a$, $m_r/H$ are large.

\subsection{What do we expect?}

For any $N \geq 2$, we can follow the dynamics of the theory described
by \Eq{Lconformal} with $\mac^2 = \tauo^{-2} (\tau/\tauo)^{n+2}$.
Here $\tauo$ is the scale where $\mac(\tauo)\tauo=1$, that is, the
point where the explicit symmetry breaking first becomes relevant.
The axion production
$\nax$ scales on dimensional grounds as $\nax \propto \fa^2 \tauo$,
so we define a dimensionless measure of axion production efficiency
\begin{equation}
  \naxbar \equiv \frac{\nax}{\fa^2 \tauo} .
  \label{naxbar}
\end{equation}
What can we say intuitively about the expected behavior for this quantity?
\begin{itemize}
\item
  For the case where $n$ is small, so the explicit symmetry breaking
  turns on gradually, we might expect that fluctuations in the fields
  lead to extra axion production.  The more field components, the more
  fluctuations there are in the fields, and so the large-$N$ case
  should intuitively produce more axions than for small $N$.
\item
  For the case where $n$ is large, the axion mass turns on rather
  abruptly.  In this case, the field rather suddenly finds itself with
  a large mass.  Locally the mass may become larger than the field's
  inverse coherence length -- that is, the potential energy from the
  tilted potential may exceed gradient energies -- and field gradients
  and gradient energies would then play little role in the subsequent
  field oscillations.  In this case we would actually expect the
  misalignment estimate to be fairly close to the actual behavior.
\item
  According to the misalignment mechanism estimate, $\naxbar$ is the
  largest for small $N$, that is, few field components.
\end{itemize}
Intuitively, then, we might expect that for small $n$, the axion
production substantially exceeds the misalignment estimate and is
larger at larger $N$, while for large $n$ (a rapid turn-on of explicit
symmetry breaking), the misalignment estimate is rather close to the
true behavior and the axion production gets \textsl{smaller} as we
consider theories with more field components.

\section{Numerical approach}
\label{sec:numerics}

Here we present our numerical implementation and our extraction of
continuum results from lattice calculations.

\subsection{Lattice implementation}

For numerical simulations, we need to discretise the equations of
motion that result from Equation \eqref{Lconformal} and express them
in terms of suitable dimensionless quantities.
The equation of motion for the
field $\Phi_j$, $j=1,\ldots,N$, is
\be
\label{eqnConfTime2Res1}
0 =  \left[ \partial_\tau^2 +  \frac{2}{\tau} \partial_\tau
  - \grad^2 + \left(\frac{\tau}{\tau_0}\right)^2
  \frac{\lambda}{2}\left(\sum_i\Phi_i^2-f_a^2\right) \right]\Phi_{j}
-\mac^2 \fa\delta_{j1}\,.
\ee
For easier notation
and to make the fields dimensionless, we rescale $\Phi_i \longrightarrow \Phi_i/\fa$,
which corresponds to scaling the vacuum manifold to the unit $N-1$-Sphere, additionally
requiring the redefinition $\lambda \longrightarrow \lambda\fa^2$. For the numerical
evolution of the fields to be stable for a significant dynamic range, one leaves out
the extra scaling of $\tau^2$ in front of $\lambda$. This means that the radial
mass is then not fixed in ``physical'' mass but rather in conformal mass, thus fixing
the size of monopole cores in terms of lattice units rather than accounting for them
growing smaller due to hubble expansion until the lattice could not resolve these
defects any more.  This is sensible if the goal is to understand the
behavior when the core size is very small, and in particular we want
in the end to extrapolate to the limit $m_r \tauo \gg 1$, which is
more easily accomplished with this treatment.  For the axion mass, on
the other hand, the $\tau^2$ scaling is physically relevant and we
keep it:
\be
\mac^2 = m_a^2 \frac{\tau^2}{\tauo^2} = \frac{1}{\tauo^2}
\left( \frac{\tau}{\tauo} \right)^{n+2} \qquad \mbox{and} \qquad
\lambda \equiv m_r^2 \,.
\ee
We shall investigate the ``axion-like'' case of $n = 7$ and compare it
to the constant physical mass case of $n = 0$.  We lattice discretize
the equation of motion with a leapfrog scheme, except that we use
backwards differences for the single $\tau$-derivative.  This is still
consistent with an $a^2$-accurate algorithm because this term is
suppressed by a $1/\tau$ leading coefficient.  Explicitly, our update
rule is:
\begin{align}
\begin{split}
  \Phi_i(\tau+\tStep,x) =&\, 2\Phi_i(\tau,x)-\Phi(\tau-\tStep,x)
  - \frac{2\tStep}{\tau}\left(\Phi_i(\tau,x)-\Phi_i(\tau-\tStep,x) \right)\\
  &+\tStepsq\sum_{j=1}^{3} \frac{\Phi_i(\tau,x + \xStep_j)
    +\Phi_i(\tau,x-\xStep_j)-2\Phi_i(\tau,x)}{\xStepsq}\\
&-
  \tStepsq\xStepsq\frac{m_r^2}{2}\left(\sum_i\Phi_i(\tau,x)^2-1\right)\Phi_i
  + \delta_{j1} \frac{\tStepsq}{\tauo^2} \left(\frac{\tau}{\tauo}\right)^{n+2}.
\end{split}
\label{updateAlgo}
\end{align}
Because $\Phi(\tau-\tStep)$ is only used locally, we can directly replace
$\Phi(\tau-\tStep)$ with $\Phi(\tau+\tStep)$ in memory, so our
memory footprint is the field values on two time slices.
We simulate on a cubic box of size
$L = N_x\xStep$ with $N_x$ some integer and impose periodic boundary conditions.
This emulates behavior on an unbounded volume as long as $\tau < \frac{L}{2}$,
which is when the lightlike signals from an event near $\tau=0$ can
first encounter each other around the periodicity.  If the speed of
information propagation is lower than the speed of light, then this
threshhold is respectively increased. The implementation is written
in C++ using AVX512 intrinsics and OMP for parallelisation. Initial
values are generated by randomly sampling an $N$-tuple from a unit
gaussian distribution and rescaling to the vacuum manifold at each
lattice site. For pseudo-random number generation we use the PCG
Random C++ header library, which in particular is well suited for
providing multiple streams of random numbers for parallel use.
Using the update formula we can now numerically evolve the axion fields and
then subsequently extract the axion number.  We do this by
implementing \Eq{naxdef} and \Eq{naxbar}, using the FFTW package for C++ as an
efficient implementation of the Fourier transform.  To optimize the
Fourier transform efficiency, we work on cubic lattices with
$N_x = 1024$, 1536, and 2048.  Since the particle number depends both
on $\Phi$ and on $\dot\Phi$, we compute $\dot\Phi$ using the
difference between two time slices; the $\Phi$-dependent part is
computed on each time slice and averaged.

We can also use our code to evaluate $\naxbar$ in the misalignment
approximation, simply by leaving out the middle line of
\Eq{updateAlgo}, which removes the gradient term and turns the code into
an independent evolution at each lattice site.  Then we also need to
replace $\sqrt{k^2+m^2} \to m$ in \Eq{naxdef}.  We checked that this
results in the same $\naxbar$ result as we get by using an adaptive
differential equation solver to evaluate \Eq{eqn:mislagnOde},
evaluating $\nax(\tA)$, and numerically integrating \Eq{naxmisalign}.

\subsection{Late-time, large-mass, continuum extrapolations}

The previous subsection shows how to numerically evaluate $\naxbar$ for a
given number of field components $N$ and mass evolution $n$, at a
given value of three lattice parameters;
the final time $\tau/\tauo$, the radial mass in physical units
$m_r \tauo$, and the lattice spacing in terms of the radial mass
$m_r \xStep$.  Of these, $N$ and $n$ represent distinct physical
problems which we want to understand.  But the other three parameters
are nuissance lattice parameters whose influences must be extrapolated
away; the desired physical result requires the joint
limits $\tau/\tauo \to \infty$, $m_r \tauo \to \infty$, and
$m_r \xStep \to 0$.  Unfortunately the product
$\tau/\xStep = (\tau/\tauo)(\tauo m_r)/(m_r \xStep)$ is bounded by $N_x$,
so it is impossible to simultaneously take all three limits, and we
will have to perform a careful extrapolation.

\begin{figure}[th]
  \includegraphics[width=\textwidth]{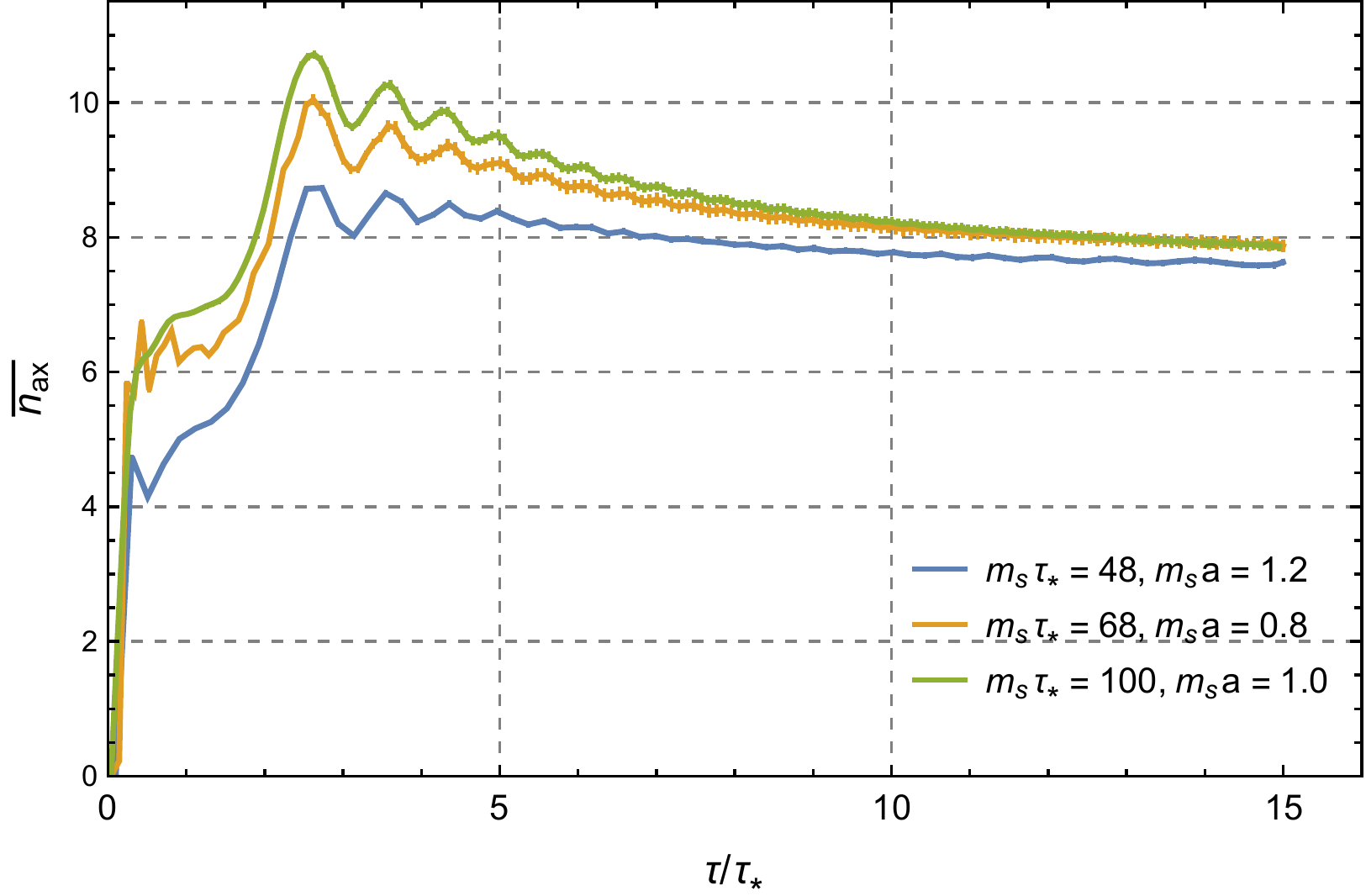}
  \caption{Axion production efficiency $\naxbar$ for fixed $N=3,n=0$
     and several $m_r \xStep$, $m_r \tauo$ values.
    After complex early dynamics, there is a slow decay superposed
    with oscillations at a fequency of $\omega = 2m_a$.}
  \label{fig:constantMassData}
\end{figure}  

First consider $\tau/\tauo$.  We can make this quantity
large by simply running our evolution for a long time.  The longer we
run, the larger $m_a$ becomes; information therefore propagates more
slowly, and the requirement $\tau/\xStep < N_x/2$ need not be strictly
enforced.  But at the same time, the ratio $m_a/m_r$ increases.
When this ratio reaches $1/2$, a process in which
two angular excitations merge into a radial excitation, which is
unphysical, becomes efficient, which can deplete the generated
axion number, changing our results.  Therefore we are obliged to terminate
our evolution while $m_a < 0.4 m_r$ to avoid this.  However, we find
that, after complex early dynamics, $\naxbar$ rather rapidly
approaches its late-time behavior.  The worst case occurs when
the physical mass is fixed; the time evolution of $\naxbar$ for
a range of other lattice parameters in the case of $n=0$, $N=3$ is
shown in Figure \ref{fig:constantMassData}.  After complex and
interesting early dynamics, the $\naxbar$ value settles towards a
large-value asymptote, plus oscillations and an inverse-power tail.
The oscillations represent some coherence in the fluctuations about
$\tA=0$ and have a frequency of $2m_a$; we eliminate them by always
evaluating $\naxbar$ at two times separated by $\pi/2m_a$ and
averaging.  This leaves a power-law decay towards the asymptotic
value.  The difference from the asymptotic value represents
finite-angle effects and decays as
$\sim \sum_{a\neq 1} \langle \Phi_a^2\rangle \propto (\tauo/\tau)^{3+n/2}$.
This functional form gives a good fit to the late-time behavior
and we use it to extrapolate $(\tau/\tauo) \to \infty$ over the
range $(\tau/\tauo) \in [10,15]$ for $n=0$ and
$(\tau/\tauo) \in [2.4,2.8]$ for $n=7$. We do not perform this
fitting procedure for $N=3$ and $n=7$ and instead just average
two evaluations at $(\tau/\tauo)=2.8$ and $\pi/2\ma$ later, becasue the 
output of simulations was initially not set up for the extrapolation procedure.
This does not significantly influence results though, because for $n=7$ the 
inverse power tail decays very rapidly. We verify  on $N=4$ and $N=5$
 that this method agrees with the extrapolation results to within 1\%,
where both methods are applicable on the more verbose output of those 
simulations.

\begin{table}[th]
\label{tab:constantMassValues3}
	\begin{tabular}{c|c|c|c|c}
	  \backslashbox{$m_r \xStep$}{$\tauo/\xStep$}
          & $40$ & $50$ &
         $85$&  $100$ \\ \hline
	$0.8$ & $7.12 \pm 0.03$ &$7.37 \pm 0.03$ &$7.72\pm 0.04$& $-$ \\\cline{2-5}
	$1.0$& $7.44 \pm 0.04$ & $7.69 \pm 0.07$& $7.87 \pm 0.09$& $7.76 \pm 0.05$ \\\cline{2-5}
	$1.2$& $7.59\pm0.03$&$7.68 \pm 0.02$& $-$ & $7.85 \pm 0.04$
	\end{tabular}
\caption{Extracted values for $\naxbar$ for $N=3$, $n=0$ and several
  other parameter values.}
\end{table}

\begin{table}[th]
\label{tab:fastMassValues3}
	\begin{tabular}{c| c| c| c| c}
	 \backslashbox{$m_r\xStep$}{$\tauo/\xStep$} & $230$ & $270$ &
         $320$&  $360$ \\ \hline
	$0.8$ & $ - $ &$ - $ &$10.58 \pm 0.18$& $10.86 \pm 0.27$ \\\cline{2-5}
	$1.0$& $ - $ & $10.69 \pm 0.26$& $10.81 \pm 0.15$& $10.75 \pm 0.17$ \\\cline{2-5}
	$1.2$& $10.45\pm 0.20$&$10.68\pm 0.31$& $10.97\pm 0.18$ & $10.74 \pm 0.20$
	\end{tabular}
\caption{Extracted $\naxbar$ values for $N=3$ fields with $n=7$
  (rapidly increasing explicit symmetry breaking) for several other
  parameter values.}
\end{table}

The resulting $\naxbar$ is displayed for $N=3$ field components
and $n=0$ (fixed physical angular mass) for several combinations
of $(m_r \tauo,m_r \xStep)$ in Table \ref{tab:constantMassValues3},
and for the $n=7$ case (rapidly increasing angular mass)
in Table \ref{tab:fastMassValues3}.
We need to perform an extrapolation of this data, and similar data for
other $N,n$ values, to the $m_r \tauo \to \infty$ and
$m_r \xStep \to 0$ limits.  To do so it is necessary to establish the
expected functional forms for each parameter dependence.
We expect that the most infrared wave number relevant to the problem
should be $k \sim \tauo^{-1}$, the coherence length of the field at
the time when the axion mass becomes important.  The most UV scale is
$k \sim m_r$, beyond which the field is no longer constrained to lie
on the vacuum manifold.  Provided that $m_r \xStep \ll \pi$, our
lattice should be able to resolve physics on this most UV scale with
errors which scale as $(m_r \xStep)^2$, since our nearest-neighbor
equations of motion and quadratic time update receive corrections of
this order.  Therefore we will assume lattice-spacing effects of this
functional form.

To determine the dependence on $(m_r \tauo)$, we need to determine
which scales are actually important to the final axion number count.
If \textsl{only} the scale $k \sim \tauo^{-1}$ is relevant and in the
absence of any topological objects, the only role of $m_r$ would be to
keep the field on the vacuum manifold.  Since finite $m_r$ does not do
so perfectly, the field could move slightly off its manifold by an
amount $\sim (k |\Phi|/m_r)^2 \sim (m_r \tauo)^{-2}$, and we would
expect corrections of this order.
However, the early-time dynamics of the fields should enter a scaling
regime which, we expect, generates a spectrum of fluctuations with all
$k$ up to $k \sim m_r$.  Typically such scaling dynamics generates
fluctuations with equal energy per logarithmic wave-number interval,
$\varepsilon \propto \int dk/k$, which according to \Eq{naxdef} means
$\nax \propto \int dk/k^2$.  Since $m_r$ acts as an artificial UV
cutoff on this integration, we expect an $\mathcal{O}(1/m_r \tauo)$
sized artifact due to finite $m_r$.  Therefore the
extrapolation to the ``stiff'' $m_r \to \infty$ limit should involve
an inverse-linear power of $m_r \tauo$.

\begin{figure}
  \includegraphics[width=0.9\textwidth]{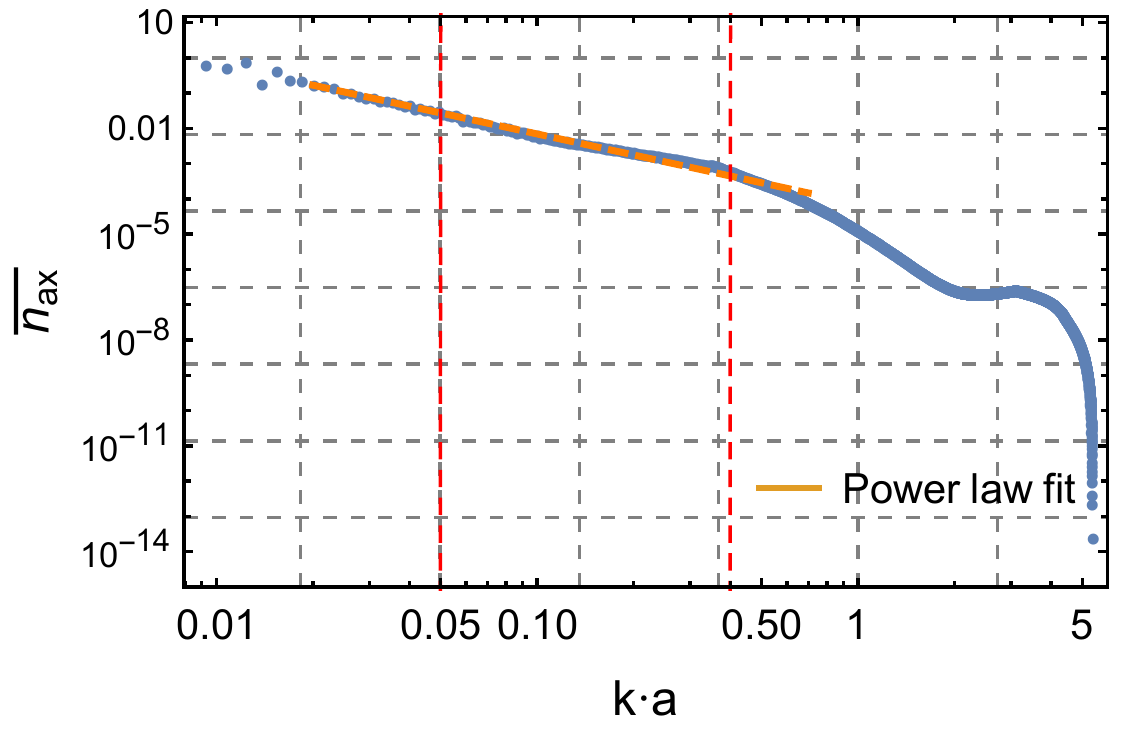}
  \caption{Log-log plot of the Fourier spectrum of the axion number
    $d\nax/dk$, for a four-component field with rapid turn-on of the
    explicit symmetry-breaking mass $n=7$, with $m_r \xStep=1.0$ and
    $m_r \tauo = 320$. Dashed red lines indicate the fitting region for
    a power law dependence. In this region, the axion number scales as
    $d\nax \propto dk/k^2$, indicating a linear sensitivity to the UV cutoff
    at the scale $m_r$. The behavior above $ka=0.5$ is modified by
    radial dynamics and the behavior above $ka=2$ is affected by the
    lattice regularization, but neither region contributes significantly.
    \label{fig:4SpectrumFastFit}
  }
\end{figure}

We can check whether this is the case by computing the spectrum of
axion fluctuations numerically from our final-time field
configurations; we simply don't do the $k$ integral in \Eq{naxdef} but
instead examine the Fourier spectrum of $\nax$.  The result for the
representative value $N=4,n=7$ is shown in Figure \ref{fig:4SpectrumFastFit},
which shows
that the spectrum of excitations decays at large $k$ in the expected
way. In particular, a power law $A\cdot x^b$ is fit to the region of the spectrum indicated,
giving $b = -1.98(2)$ for Figure \ref{fig:4SpectrumFastFit}.
As such, an $\mathcal{O}(1/m_r \tauo)$ fraction of the axion
number should reside beyond the correctly-sampled scales and we
therefore require $1/m_r \tauo$ corrections in our extrapolation. 
Note that the presence of monopole topological defects could also lead
to $1/m_r \tauo$ corrections, since an $\mathcal{O}(1/m_r)$ fraction
of the monopole energy is contained in the core where the field
unphysically departs from the vacuum manifold.

What about the lattice spacing?  For $ka \ll 1$ the lattice treatment
we use is accurate up to $(ka)^2$ corrections.  This is largest for
$k \sim m_r$, the shortest physically relevant scale.  At this scale,
lattice-spacing effects give rise to $(m_r a)^2$-suppressed
corrections.  But we have just seen that this scale only gives rise to
an $\mathcal{O}(m_r \tauo)$-suppressed fraction of the axion number;
therefore the expected size of lattice-spacing artifacts in the
determined $\naxbar$ value is of order $m_r a^2/\tauo$.
So the functional form we use to extrapolate to the small $a$
and large $m_r$ limits is
\begin{equation}
  \naxbar(m_r \tauo,m_r a) = (\naxbar)_c + A (m_r \tauo)^{-1}
  + B m_r a^2 / \tauo
\end{equation}
where $ (\naxbar)_c$ is the desired continuum limit and $A,B$ are
corrections.  We find that this fitting form gives a good
representation of our data, with
$\chi^2/\mathrm{dof}$ of $2.114$ when $n=0$ and $0.404$ when $ n=7$,
 respectively.

\begin{table}[th]

	\begin{tabular}{c|c|c|c|c}
	  \backslashbox{$m_r \xStep$}{$\tauo/\xStep$}
          & $40$ & $50$ &
         $85$&  $100$ \\ \hline
	$0.8$ & $6.50 \pm 0.03$ &$6.78 \pm 0.03$ &$7.05\pm 0.08$& $-$ \\\cline{2-5}
	$1.0$& $6.74 \pm 0.03$ & $6.93 \pm 0.02$& $7.19 \pm 0.04$& $7.20 \pm 0.05$ \\\cline{2-5}
	$1.2$& $6.84\pm0.04$&$7.03 \pm 0.03$& $-$ & $7.26 \pm 0.05$
	\end{tabular}
\caption{Extracted values for $\naxbar$ for $N=4$, $n=0$ and several
  other parameter values.}
\label{tab:constantMassValues4}
\end{table}

\begin{table}[th]
	\begin{tabular}{c| c| c| c| c}
	 \backslashbox{$m_r\xStep$}{$\tauo/\xStep$} & $230$ & $270$ &
         $320$&  $360$ \\ \hline
	$0.8$ & $ - $ &$ - $ &$11.27 \pm 0.21$& $10.79 \pm 0.29$ \\\cline{2-5}
	$1.0$& $ 10.72\pm0.23 $ & $10.20 \pm 0.40$& $10.39 \pm 0.30$& $10.70 \pm 0.4$ \\\cline{2-5}
	$1.2$& $10.28\pm0.15$&$10.60\pm 0.22$& $10.50\pm 0.40$ & $10.48 \pm 0.25$
	\end{tabular}
\caption{Extracted $\naxbar$ values for $N=4$ fields with $n=7$
  (rapidly increasing explicit symmetry breaking) for several other
  parameter values.}
\label{tab:fastMassValues4}
\end{table}

\begin{table}[th]
	\begin{tabular}{c|c|c|c|c}
	  \backslashbox{$m_r \xStep$}{$\tauo/\xStep$}
                     & $40$ & $50$ & $85$&  $100$ \\ \hline
	$0.8$ & $6.06\pm0.03$ &$6.26\pm0.05$ &$6.57\pm0.05$& $-$ \\\cline{2-5}
	$1.0$& $6.26\pm0.03$ & $6.42\pm0.03$& $6.50\pm0.03$& $6.69 \pm 0.03$ \\\cline{2-5}
	$1.2$& $6.33\pm0.03$ & $6.46\pm0.03$& $-$ & $6.51\pm0.04$
	\end{tabular}
\caption{Extracted values for $\naxbar$ for $N=5$, $n=0$ and several
  other parameter values.}
\label{tab:constantMassValues5}
\end{table}

\begin{table}[th]
	\begin{tabular}{c| c| c| c| c}
	 \backslashbox{$m_r\xStep$}{$\tauo/\xStep$} & $230$ & $270$ &
         $320$&  $360$ \\ \hline
	$0.8$ & $ - $ &$ - $ &$9.8\pm 0.5$& $8.7 \pm 0.6$ \\\cline{2-5}
	$1.0$& $ - $ & $10.2\pm0.4$& $11.0 \pm 0.5$& $9.7 \pm 0.4$ \\\cline{2-5}
	$1.2$& $10.06 \pm 0.26$&$9.7\pm0.4$& $10.5 \pm 0.7$ & $10.1\pm 0.4$
	\end{tabular}
\caption{Extracted $\naxbar$ values for $N=5$ fields with $n=7$
  (rapidly increasing explicit symmetry breaking) for several other
  parameter values.}
\label{tab:fastMassValues5}
\end{table}
We can repeat this extrapolation process now for N=4 and N=5.
The results of this are shown in Tables \ref{tab:constantMassValues4}, \ref{tab:fastMassValues4}, \ref{tab:constantMassValues5}, \ref{tab:fastMassValues5}.
For $N=4$ one finds a $\chi^2/\mathrm{dof}$ of $1.0853$ and $0.7354$ when $n=0$ or $n=7$. For $N=5$ one has $2.8566$ and $1.7671$ respectively.

\section{Results}
\label{sec:results}

In the previous section we saw how to carry out extrapolations of
finite-spacing, finite-$m_r$ data to the continuum and
heavy-radial-field limits.  We have done so for the cases of $N=3,4,5$
field components for each case of interest here -- a fixed physical
mass $n=0$ (that is, $\mac^2 = \tau^2/\tauo^4$) and a rapidly growing
mass $n=7$ or $\mac^2 = \tau^9/\tauo^{11}$, analogous to how the axion mass
increases cosmologically.  Our central results are the axion
production efficiencies for each of these cases, extrapolated to the
late-time, fine-spacing, and large-radial-mass limits, which we
present in Table \ref{tab:mainresults}.  Each result in the table
represents an extrapolation over 10 lattice spacing/$\tauo$
combinations, each of which involved averaging at least $10$
independent lattice evolutions when $n=7$ and at least $6$ when $n = 0$,
except when $N=5$. Because $N=5$ is computationally more expensive, 
only half as many independent evolutions were performed, leading to increased
statistical errors.
  In every case the continuum and
large-$m_r$ extrapolations are very mild, with the final result
always differing by less than 10\% from the coarsest and
smallest-$m_r$ lattice.  The table represents the main results of
this work.

\begin{table}
  \centerline{
  \begin{tabular}{|c|c|c|r|c|} \hline
    $N$ & $n$ & $\naxbar$ & $\nmissbar$ & ratio \\ \hline
    3 & 0 & $8.10\pm0.06$ &  5.47 & $1.48\pm 0.01$ \\
    4 & 0 & $7.58\pm0.05$ &  4.58 & $1.65\pm 0.01$ \\
    5 & 0 & $6.78\pm0.06$ &  4.18 & $1.62\pm 0.02$\\ \hline
    3 & 7 & $11.52\pm0.30$ & 10.91 & $1.05\pm0.03$ \\
    4 & 7 & $10.66\pm0.44$ &  9.47 & $1.12\pm0.05$ \\
    5 & 7 & $10.05\pm1.20$ &  8.82 & $ 1.14\pm0.14$\\ \hline
  \end{tabular}  }
  \caption{Dimensionless ``axion'' production efficiency (middle
    column), misalignment expectation, and their ratio (last column)
    for three values of $N$ the number of field components and two
    rates of angular-excitation mass-growth $n$.}
  \label{tab:mainresults}
\end{table}

\section{Discussion and conclusions}
\label{sec:conclusions}

As we see in the previous section, the production efficiency for
``axions'' is significantly higher than the misalignment estimate
\textsl{if} the axion mass turns on gradually -- for instance, if it
remains fixed in physical units.  In this case, the chaotic spatial
distribution of the field leads to additional fluctuations.
The discrepancy appears to get larger with more independent field
components, but this turns out to be a weak effect; the total
production declines with increasing $N$, at least for $N=3,4,5$.

However, when the axion mass turns on more abruptly -- for instance,
if $m_a^2 \propto T^{-7}$, close to the physically relevant case for
real axions -- then our results are quite different.  The production
of angular fluctuations is quite close to the expectation based on the
misalignment picture, and the expected trend -- that the production is
smaller for fields with more components -- does emerge.

Naturally one cannot immediately take this result and conclude that
the same is true of the $N=2$ field-component case, that is, actual
axions.  The presence of string topological defects, and their large
role in the energy budget, is a significant difference from these
higher-field-component models.  Nevertheless, we have learned
something important.  When the explicit symmetry breaking turns on
rapidly, that significantly changes the dynamics.  The field rapidly
becoming heavy makes gradient terms and long-range variation less
important; the fields' evolution become much more determined by the
local field value and less by spatially varying structures.  This
could at least partly explain why the production efficiency for axions
is also surprisingly small in the case where the axion mass grows
as a high power of the conformal time.

\bibliographystyle{unsrt}
\bibliography{refs}

\end{document}